\documentclass[twocolumn,showpacs,fleqn,nobibnotes]{revtex4}

\usepackage{amsmath}
\usepackage{graphicx}
\usepackage{float}
\usepackage{subfigure}
\usepackage{hyperref}

\def\lsim{\raise0.3ex\hbox{$<$\kern-0.75em\raise-1.1ex\hbox{$\sim$}}}
\def\gsim{\raise0.3ex\hbox{$>$\kern-0.75em\raise-1.1ex\hbox{$\sim$}}}

\newcommand{\beq}{\begin{equation}}
\newcommand{\eeq}{\end{equation}}

\begin{document}

\title{Study on the PDF comparisons for quarkonium + $\gamma$ production\\ at the LHC and FCC energies}
\author{M.M. Machado}
\affiliation{Instituto Federal Farroupilha - Campus S\~ao Borja\\
CEP 97670-000, S\~ao Borja, RS, Brazil.}

\author{G. Gil da Silveira}
\affiliation{Instituto de F\'{\i}sica, Universidade Federal do Rio Grande do Sul\\
Caixa Postal 15051, CEP 91501-970, Porto Alegre, RS, Brazil}

\begin{abstract}
The quarkonium plus photon production in coherent hadron-hadron interactions at the LHC is studied using the non-relativistic QCD factorization formalism. We investigate a set of kinematic distributions and compute the total cross sections for $J/\Psi + \gamma $ production. Our results demonstrate the feasibility of such process in the LHC kinematic regime and explore the possibilities for the Future Circular Collider, where higher event yields can be achieved.
\end{abstract}
\maketitle
\section{Introduction}

Over the last years, the quarkonium production becomes the subject of intense theoretical and experimental investigations. This interest has been motivated by the observation of
large discrepancies between experimental measurements of $J/\Psi$ production at the Collider Detector (CDF) at Fermilab \cite{CDF} and theoretical calculations based on the color-singlet model (CSM) \cite{csm}. Attempts to understand this discrepancy focus on new production mechanisms that allow the $c\bar{c}$ bound system to be produced in a color-octet state and evolve non-perturbatively into a charmonium, where its large mass provides a natural hard scale that allows the application of perturbative Quantum Chromodynamics (QCD). There are several mechanisms proposed for the quarkonium production at hadron colliders \cite{Lansberg}, as the CSM, the color-octet model (COM) \cite{com}, and the color evaporation model (CEM) \cite{cem}. Also, the diffractive quarkonium production is sensitive to the gluon content of the Pomeron at small-$x$ \cite{forward} and may be particularly useful in studying the different mechanisms for quarkonium production. Hence, considering the experimental point-of-view, the heavy-quarkonium photoproduction have an extremely clean signature through their leptonic decay modes \cite{forward}. 

Studies of $\gamma$-proton interactions at the Large Hadron Collider (LHC) provides valuable information on the QCD dynamics at high energies. The photon–hadron interactions can be divided into {\it{exclusive}} and {\it{inclusive}} reactions. In the first case, a certain particle is produced, while the target remains in the ground state (or is internally excited only). On the other hand, in inclusive interactions the particle produced is accompanied by one or more particles from the breakup of the target.

In this work, we are focused in the process
\begin{equation}
\label{process}
p + p \rightarrow p \oplus J/\Psi + \gamma + X.
\end{equation}
i.e., the inclusive quarkonium + photon production in $pp$ collisions at the LHC and Future Circular Collider (FCC) energies, and predict their cross sections as a function of the quarkonium rapidity ($y$) and transverse momentum ($p_{\perp}$). The $\oplus$ represent the presence of rapidity gaps between the colliding proton and the produced meson. Such processes are relatively easy to be detected through their leptonic decay modes, and their transverse momenta are balanced by the associated high-energy photon. While the LHC is colliding protons at staggering 13~TeV, the FCC is planned to collide protons at 100~TeV~\cite{Mangano}, and will be an important step for the future development of high energies physics. Our predictions are the first one for the energy scale of this collider.

In next section we present a brief review about the quarkonium$+\gamma$ photoproduction in the nonrelativistic QCD (NRQCD) formalism. Next, we present our predictions for the rapidity and transverse momentum distributions as well as the total cross sections for $J/\Psi + \gamma$ and $\Upsilon + \gamma$ production at LHC and FCC energies.
Finally, the last section summarizes our main conclusions.

\section{The $J/\psi$ + photon photoproduction}

The total cross section for the quarkonium + photon process, $p + p \to p + M + \gamma + X $, is given by \cite{mehen} 
\begin{eqnarray}
\sigma (W^{2}_{\gamma h}) = 2 \int d\omega \frac{dN(\omega)}{d\omega}\sigma_{\gamma p\rightarrow M+\gamma +X}(W^{2}_{\gamma h}),
\label{convol}
\end{eqnarray}
where $\omega$ is the photon energy, $dN(\omega)/d\omega$ is the equivalent photon spectrum, and $W^{2}_{\gamma h} = 2\omega \sqrt{S_{NN}}$ is the center-of-mass energy in the photon-hadron system, with $\sqrt{S_{NN}}$ for the hadron-hadron system. In this work we consider the LHC energy of 13 TeV and the FCC at 100 TeV.

In the NRQCD formalism, the cross section for the production of a heavy quarkonium state $M$ factorizes into
\begin{eqnarray}
\sigma (ab \rightarrow M + X) = \Sigma_{n} \sigma(ab \rightarrow Q\bar{Q} [n] + X)\langle O^{M}[n]\rangle,
\end{eqnarray}
where the coefficients $\sigma(ab \rightarrow Q\bar{Q}[n] + X)$ are the short-distance cross sections for the production of the heavy-quark pair $Q\bar{Q}$ in an intermediate Fock state $n$, which does not have to be color neutral. The $\langle O^{M}[n]\rangle$ are nonperturbative long-distance matrix elements, which describe the transition of the intermediate $Q\bar{Q}$ state into the physical state $M$ via soft-gluon radiation. Currently, these elements have to be extracted from global fits to quarkonium data as performed in Ref.~\cite{review_nrqcd}. It is important to emphasize that the underlying mechanism governing the heavy quarkonium production is still subject of intense debate (for a recent review see, e.g., Refs.~\cite{magno,mmmmvt,vicmairon,vicmairon2}.

\begin{figure}[b]
\centering
\includegraphics*[scale=0.5]{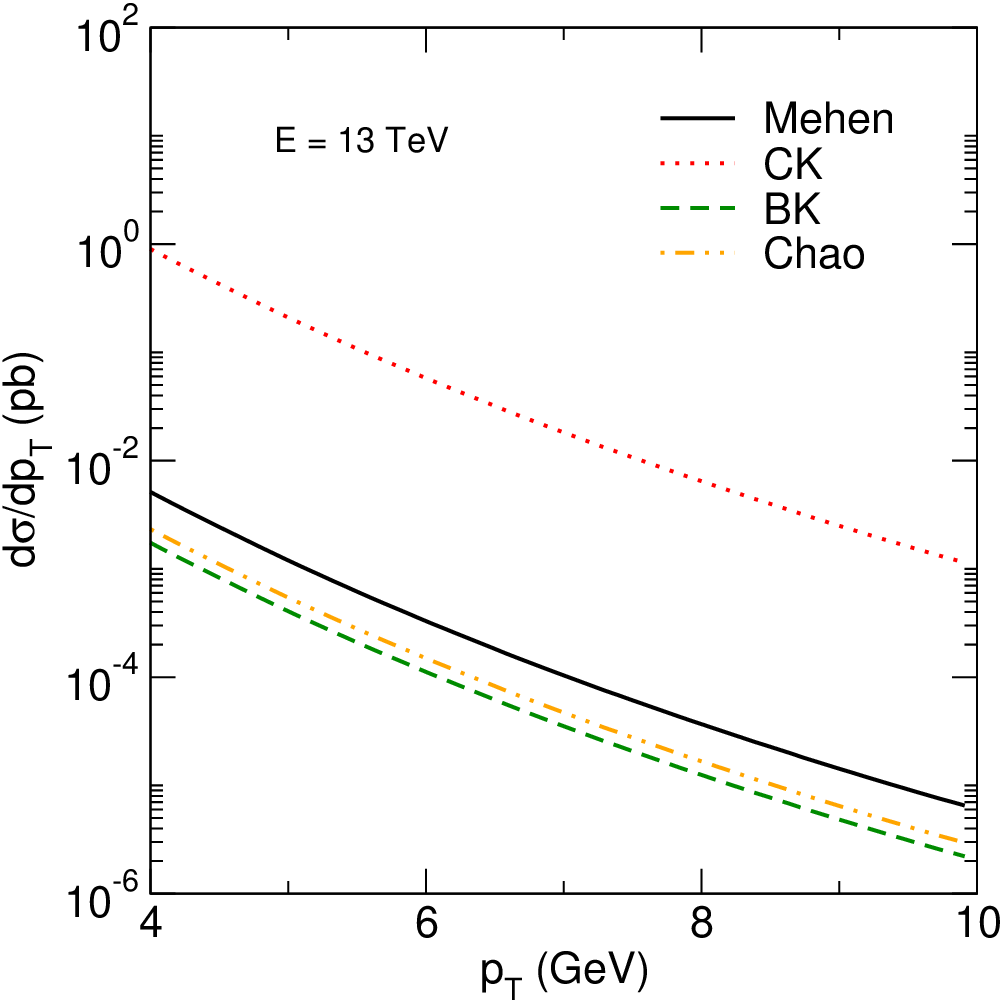} 
\caption{The $p_{T}$ distribution of the $J/\psi + \gamma$ production at 13~TeV for different matrix elements: Mehen, Cacciari \& Kramer (CK), Butenschoen \& Kniehl (BK), and Chao {\it et al.}}
\label{fig:1}
\end{figure}

In the particular case of $M + \gamma$ photoproduction, the total cross section can be expressed as follows \cite{mehen}
\begin{eqnarray}
\sigma_{\gamma p} = \int dz dp_{T}^2 \frac{xg_p(x,Q^2)}{z(1-z)} \frac{d\sigma}{dt}(\gamma + g \rightarrow M + \gamma ),
\label{sigmagamp}
\end{eqnarray}
where
\begin{eqnarray}
z = \frac{p_V\cdot p}{p_{\gamma} \cdot p},
\end{eqnarray}
with $p_V$, $p$, and $p_{\gamma}$ being the 4-momentum of the quarkonium, hadron, and photon, respectively. In the hadron rest frame, $z$ can be interpreted as the fraction of the photon energy carried away by the quarkonium.
The partonic differential cross section in Eq.~\ref{sigmagamp} is given by \cite{ko}
\begin{eqnarray}\nonumber
\frac{d\sigma}{dt} &=& \frac{64 \pi^2}{3}\frac{e_Q^4 \alpha^2\alpha_s m_Q}{s^2} \\
&\times& \left(\frac{s^2s_1^2+t^2t_1^2+u^2u_1^2}{s_1^2t_1^2u_1^2}\right) \langle O^{V}(^3S_1^{[1]})\rangle,
\label{dsdt}
\end{eqnarray}
where $e_Q$ and $m_Q$ are, respectively, the charge and mass of heavy quark constituent of the quarkonium. The Mandelstam variables can be expressed in terms of $z$ and $p_{T}$ as follows
\begin{subequations}
\begin{eqnarray}
s & = & \frac{p_{T}^2+(2m_Q)^2(1-z)}{z(1-z)}, \\
t & = & - \frac{p_{T}^2+(2m_Q)^2(1-z)}{z}, \\
u & = & - \frac{p_{T}^2}{1-z},
\end{eqnarray}
\end{subequations}
and they are combine into
\begin{subequations}
\begin{eqnarray}
s_1 = s - 4 m_Q^2, \\
t_1 = t - 4 m_Q^2, \\
u_1 = u - 4 m_Q^2,
\end{eqnarray}
\end{subequations}
with the Bjorken variable $x$ 
expressed like
\begin{eqnarray}
x = \frac{p_{T}^2+(2m_Q)^2(1-z)}{W_{\gamma h}^2z(1-z)}.
\end{eqnarray}
The long-distance matrix elements, $\langle O^{V}(^3S_1^{[1]})\rangle$, can be determined from quarkonium electromagnetic decay rates. In our calculations 	
we use the values as given in Refs.~\cite{mehen,bk,cacciari,chao} for the $J/\Psi$ production (see Tab.~\ref{tab1}). In what follows we consider different parametrizations for the parton distribution functions (PDF), in particular, we employ CT10 \cite{ct10}, CT14 \cite{ct14}, NNPDF v.~2.1 \cite{nnpdf21} and 3.1 \cite{nnpdf31}, MMHT2014 \cite{mmht}, and MSTW2008 \cite{mstw}.

\begin{table}[b]
\centering
\renewcommand{\arraystretch}{1.5}
\begin{tabular}{|l|c|c|c}
\hline\hline
$\langle O^{V}(^3S_1^{[1]})\rangle$ & $J/\Psi$ ($10^{-3}$ GeV$^{3}$) \\ \hline
Mehen \cite{mehen}    & $ 6.60 $ \\
BK    \cite{bk}       & $ 2.24 $ \\
CK    \cite{cacciari} & $ 1 600 $ \\
Chao  \cite{chao}     & $ 3.00 $ \\
\hline \hline
\end{tabular}  	
\caption{Matrix elements employed to compute the partonic differential cross section for the quarkonium photoproduction.}
\label{tab1}
\end{table}

\section{Results}

\begin{figure*}[t]
\includegraphics[scale=0.5]{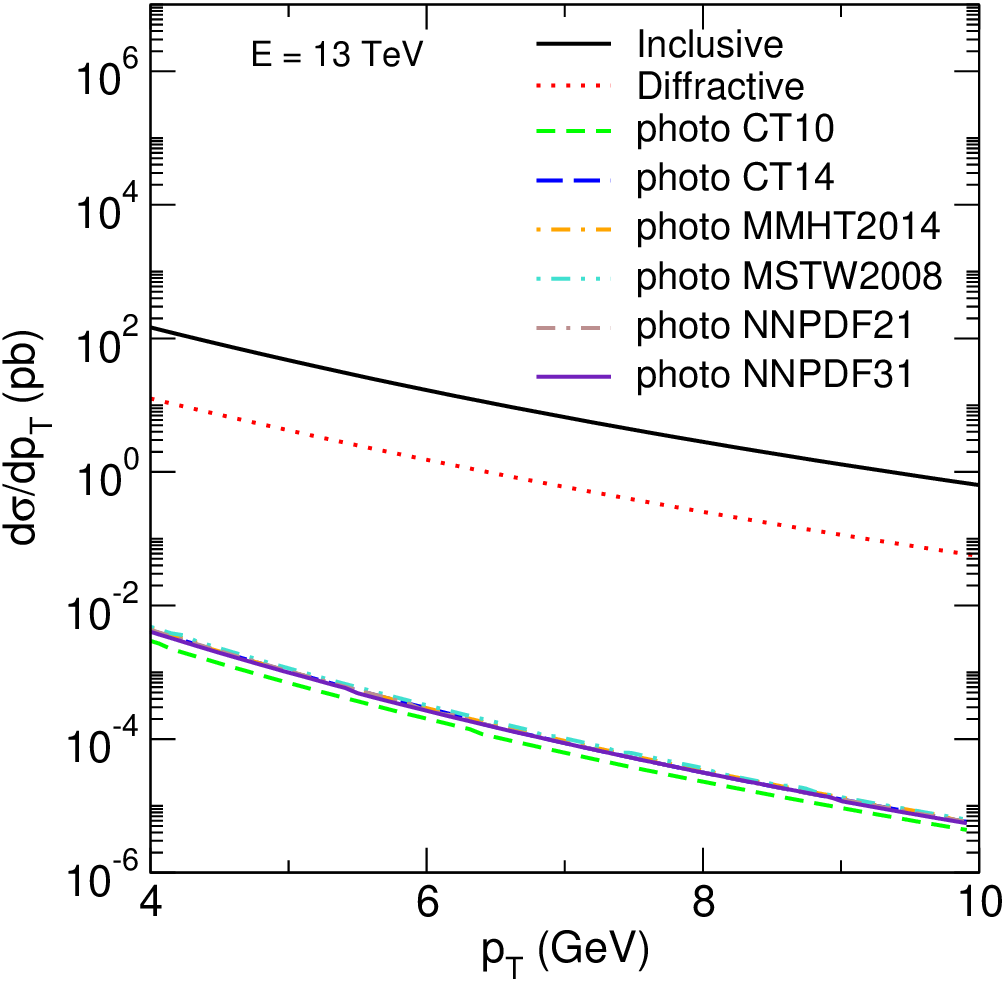} 
\includegraphics[scale=0.5]{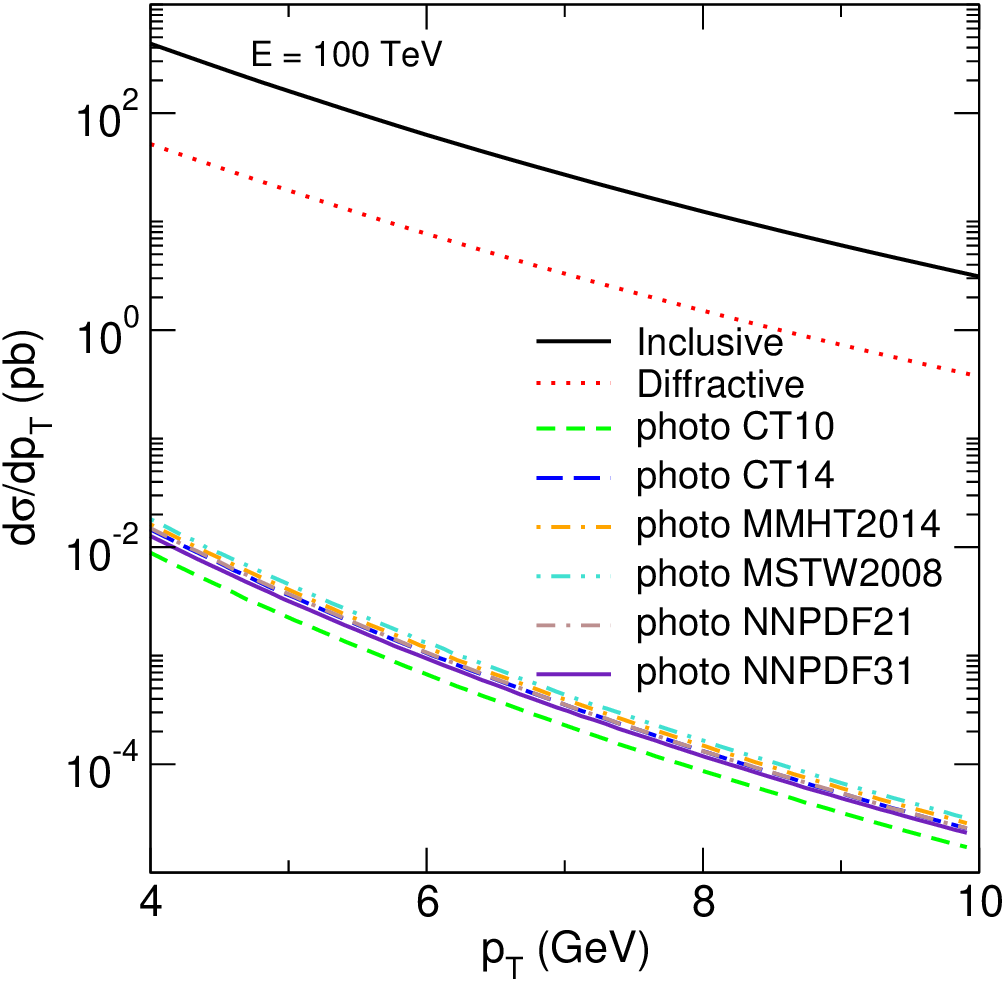} 
\caption{The $p_{T}$ distribution for 13 (top panel) and 14~TeV (bottom panel) TeV with different PDF and in comparison to the inclusive and diffractive production mechanisms.}
\label{fig:2}
\end{figure*}

Focusing in the case of $J/\psi + \gamma$ case, we compute the $p_{T}$ distribution in order to compare the different predictions among the possible matrix elements, as shown in Figs.~\ref{fig:1} and \ref{fig:2}. We also explore recent PDF parametrizations in order the evaluate the distinct contributions and possible kinematics regions to discriminate among them.

In Fig.~\ref{fig:1} we present the comparison of the different values of the NRQCD matrix elements at 13~TeV in the LHC. Although all predicions present the same slope, one clearly sees that the CK matrix element predicts a larger event rate. The CK matrix elements are the oldest estimations which are possibly superseded by the latest ones, i.e., Chao and BK.

In Fig.~\ref{fig:2}, we provide the predictions for the inclusive, diffractive, and photon-proton cross sections for different PDFs, now at 13~TeV (top panel) and 100~TeV (bottom panel). For these results we have employed the matrix element estimated by Chao {\it et al.} As a result, one can see the contribution from photoproduction is two-order of magnitude smaller than the inclusive one. Moreover, the predictions with different PDF parametrizations show a rather good agreement at the LHC energy and a slightly separation at the higher energy of FCC. In both cases, the CT10 contribution is the smallest one.

\begin{figure}[b]
\includegraphics[scale=0.55]{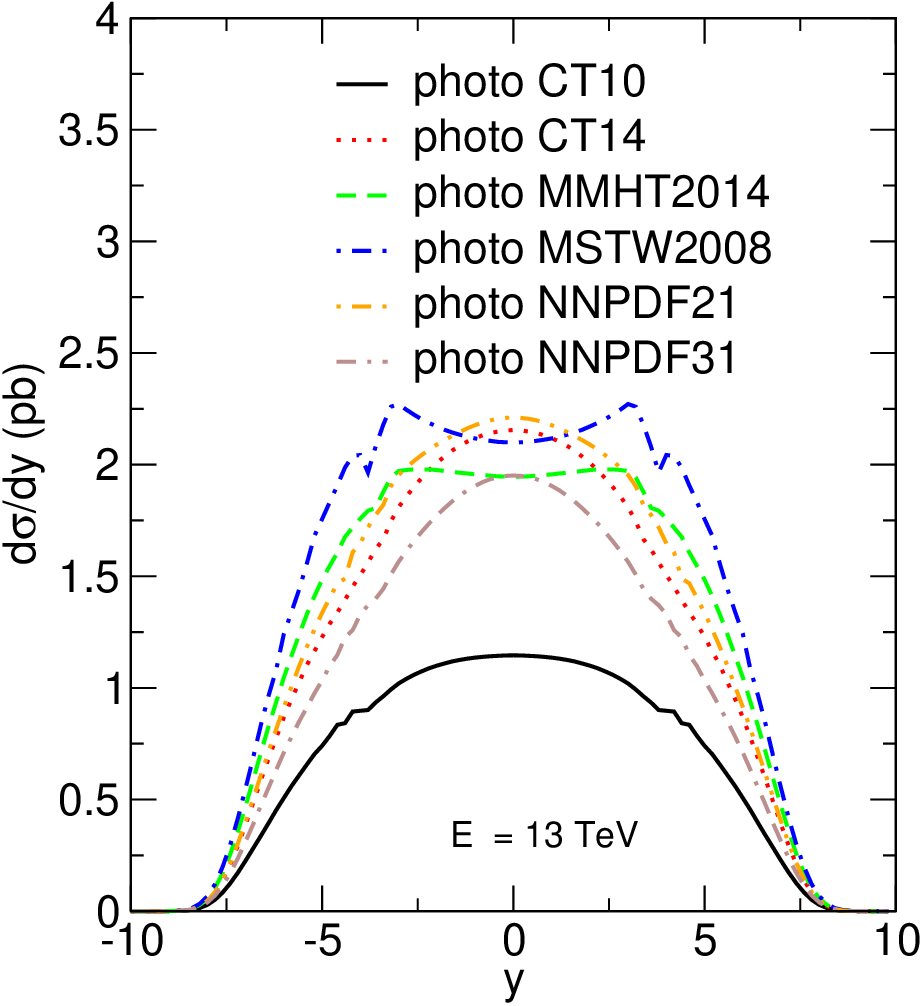} 
\caption{Comparison of the vector meson rapidity distribution for various PDF parametrizations at 13 TeV. }
\label{fig:3}
\end{figure}

We also look into the rapidity distribution in order to discriminate the different contributions. In Fig.~\ref{fig:3}, the different parametrizations are compared in the rapidity distribution of the vector meson. Although similar results are obtained, the rapidity distributions show distinct rapidity regions where each PDF have more contribution. This is an important discriminator to compare prediction with data obtained in the LHC experiments.

\begin{figure}[h]
\includegraphics[scale=0.475]{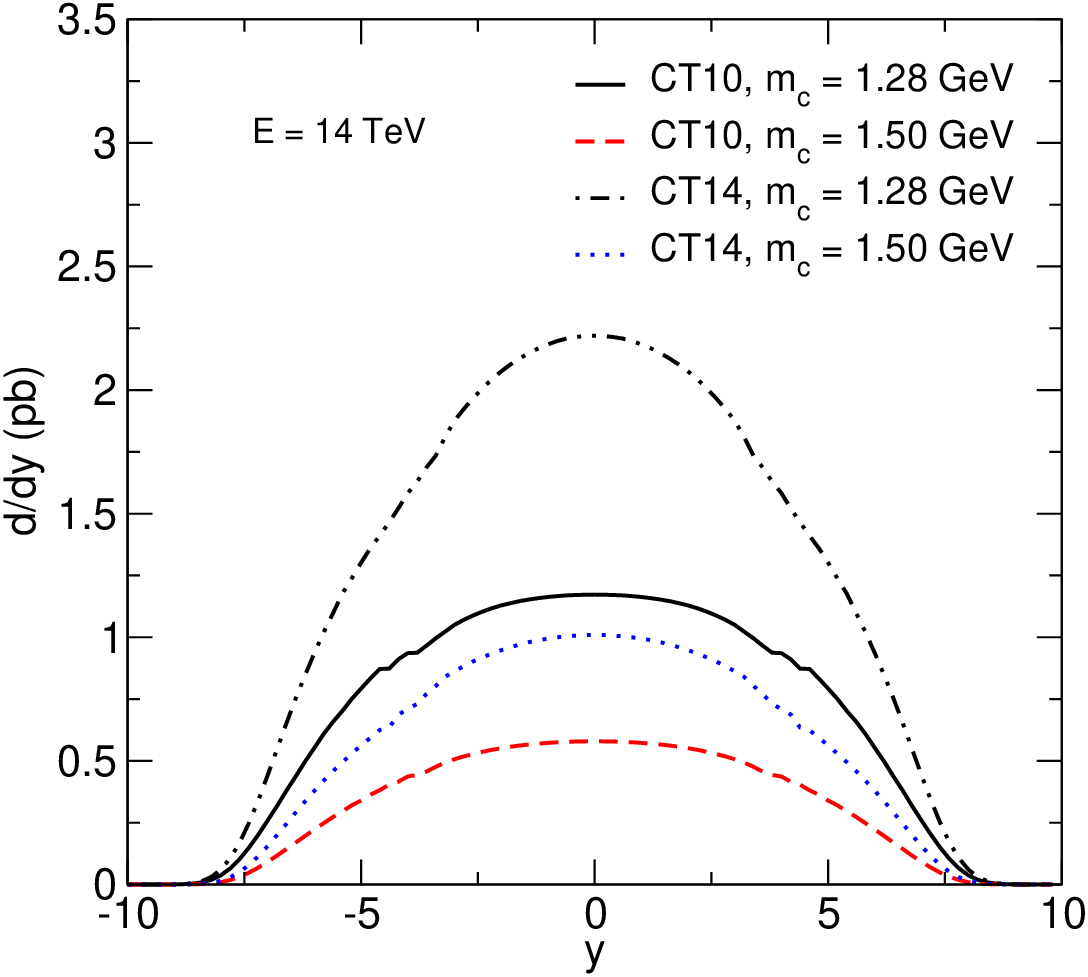} 
\caption{Rapidity distributions for the chosen charm mass values in the literatue obtained with the CT10 and CT14 PDF parametrizations. }
\label{fig:4}
\end{figure}

Another relevant result is the comparison of the chosen charm mass, as showed in Fig.~\ref{fig:4}, where the comparisons of different values of charm mass are presented. We consider two different values of charm mass: $m_{c}$~=~1.28~GeV/$c$ as found in the PDG~\cite{pdg} and the value of $m_{c}$~=~1.50~GeV/$c$ employed in several works in the literature. These predicitons are compared using two different PDF parametrizations (CT10 and CT14). As one can see in Fig.~\ref{fig:4}, there are different contributions along the central detector and more data will allow us to obtain more stringent constraints on the parametrizations.

\section{Conclusions}

In summary, we have computed the cross sections for the photoproduction of quarkonium$+ \gamma$ in coherent $pp$ collisions at LHC and FCC energies using the NRQCD formalism and considering different sets of values for the matrix elements. Such  processes are interesting since the final state is characterized by a low multiplicity and one rapidity gap. For the particular case o f $J/\psi + \gamma$ production, our results demonstrate that the $p_{T}$ and $y$ distributions are strongly dependent on the NRQCD matrix elements and the PDFs parametrization. A detailed data analysis could provide sensible results to constraint the proton PDF parametrizations and NRQCD models. An extension of this work will include a detail study on the $\Upsilon + \gamma$ production for the LHC and FCC energies.

\begin{acknowledgments}
This work was supported by CNPq, CAPES, and FAPERGS, Brazil.
\end{acknowledgments}

\end{document}